\documentclass[a4paper,fleqn]{cas-dc}

\usepackage[numbers]{natbib}

\usepackage{units}

\def\tsc#1{\csdef{#1}{\textsc{\lowercase{#1}}\xspace}}
\tsc{WGM}
\tsc{QE}
\tsc{EP}
\tsc{PMS}
\tsc{BEC}
\tsc{DE}

\begin{document}
\let\WriteBookmarks\relax
\def\floatpagepagefraction{1}
\def\textpagefraction{.001}
\shorttitle{Swelling and blistering of helium-ion-irradiated tungsten}
\shortauthors{Allen et~al.}

\title [mode = title]{Key mechanistic features of swelling and blistering of helium-ion-irradiated tungsten}                      
\author[1,2]{Frances I. Allen}[orcid=0000-0002-0311-8624]
\ead{francesallen@berkeley.edu}

\cormark[1]

\address[1]{Department of Materials Science and Engineering, UC Berkeley, Berkeley, CA 94720, USA}
\address[2]{National Center for Electron Microscopy, Molecular Foundry, LBNL, Berkeley, CA 94720, USA}

\author[3,4]{ Peter Hosemann}

\address[3]{Department of Nuclear Engineering, UC Berkeley, Berkeley, CA 94720, USA}
\address[4]{Materials Sciences Division, LBNL, Berkeley, CA 94720, USA}

\author[3]{ Mehdi Balooch}

\begin{abstract}
Helium-ion-induced swelling and blistering of single-crystal tungsten is investigated using a Helium Ion Microscope for site-specific dose-controlled irradiation (at \unit[25]{keV}) with analysis by Helium Ion Microscopy, Atomic Force Microscopy and Transmission Electron Microscopy (cross-sectioning by Focused Ion Beam milling). Our measurements show that the blister cavity forms at the depth of the helium peak and that nanobubbles coalesce to form nanocracks within the envelope of the ion stopping range, causing swelling of the blister shell. These results provide the first direct experimental evidence for the interbubble fracture mechanism proposed in the framework of the gas pressure model for blister formation.
\end{abstract}



\maketitle

Helium ion irradiation of materials can induce an evolution of complex near-surface morphology changes starting with the creation of point defects such as vacancies and interstitials, followed by diffusion and recombination processes producing gas clusters and cavities (also known as nanobubbles), finally resulting in microscopic and macroscopic changes such as fuzzing, blistering, cracking and spall-ing. An in-depth understanding of these phenomena is crucial in the development of materials destined for operation in harsh irradiation environments. For example, the plasma-facing components of nuclear fusion reactors will need to withstand high fluxes of helium ions with energies up to several thousand electronvolts, as well as hydrogen isotopes, neutrons, X-rays, and intense transient heat loads \cite{Bolt2002}.

The first to report the formation of nanobubbles in a metal irradiated with helium ions was Nelson in 1964, in experiments at room temperature and up to \unit[500]{$^\circ$C} using \unit[60]{keV} helium ions incident on copper, silver and gold \cite{Nelson1964}. At around the same time, blister formation at room temperature on copper and nickel irradiated with \unit[40--140]{keV} helium ions was also observed \cite{Primak1966}. Since then, the near-surface morphological changes induced by helium ion irradiation have been the subject of numerous investigations spanning several decades up to the present. Experimental work in this area has primarily used ion accelerators or plasma devices to achieve the irradiation conditions. These large-scale experiments have enabled a great degree of flexibility in terms of the sample type, size and temperature, yet fine control over the irradiation parameters can be challenging, in particular with regards to probing early morphological changes in a high-throughput, repeatable manner. A complimentary emerging approach in this field is to use the precise control of the Helium Ion Microscope to achieve localized dose-controlled irradiation, as demonstrated in investigations of helium nanobubble formation \cite{Wang2016,Mairov2019,Shan2019} and blistering \cite{Veligura2013,Bergner2018} for various materials, including a recent study of helium-ion-induced blistering on coarse-grained versus fine-grained tungsten \cite{Chen2018}. 

Tungsten is one of the primary candidates for the plasma-facing components in current fusion reactor design \cite{Bolt2002,Alimov2013}, hence experimental results pertaining to helium ion irradiation of tungsten targets are of particular interest. At elevated temperatures \unit[$\gtrsim$700]{$^\circ$C}, helium ion irradiation of tungsten and other metals is known to cause porous surface structures to form, instead of blisters, and it is speculated that this is due to bubble migration towards the surface \cite{Cipiti2005}. Studies of this so-called tungsten fuzz have intensified, because the temperature excursions experienced by for example the divertor of a tokamak generate the conditions for fuzz formation \cite{Petty2015,Parish2017}. While the effects of fuzz in an actual fusion reactor need to be better understood, it is also conjectured that its formation may be of benefit, since it prevents the exfoliation that accompanies blistering. In any event, blister formation in the lower temperature regimes is still a serious concern and direct experimental evidence for the underlying mechanisms of blister formation that were proposed over 40 years ago \cite{Evans1977,Evans1978} has been lacking. 

\begin{figure*}
    \centering
        \includegraphics[scale=.9]{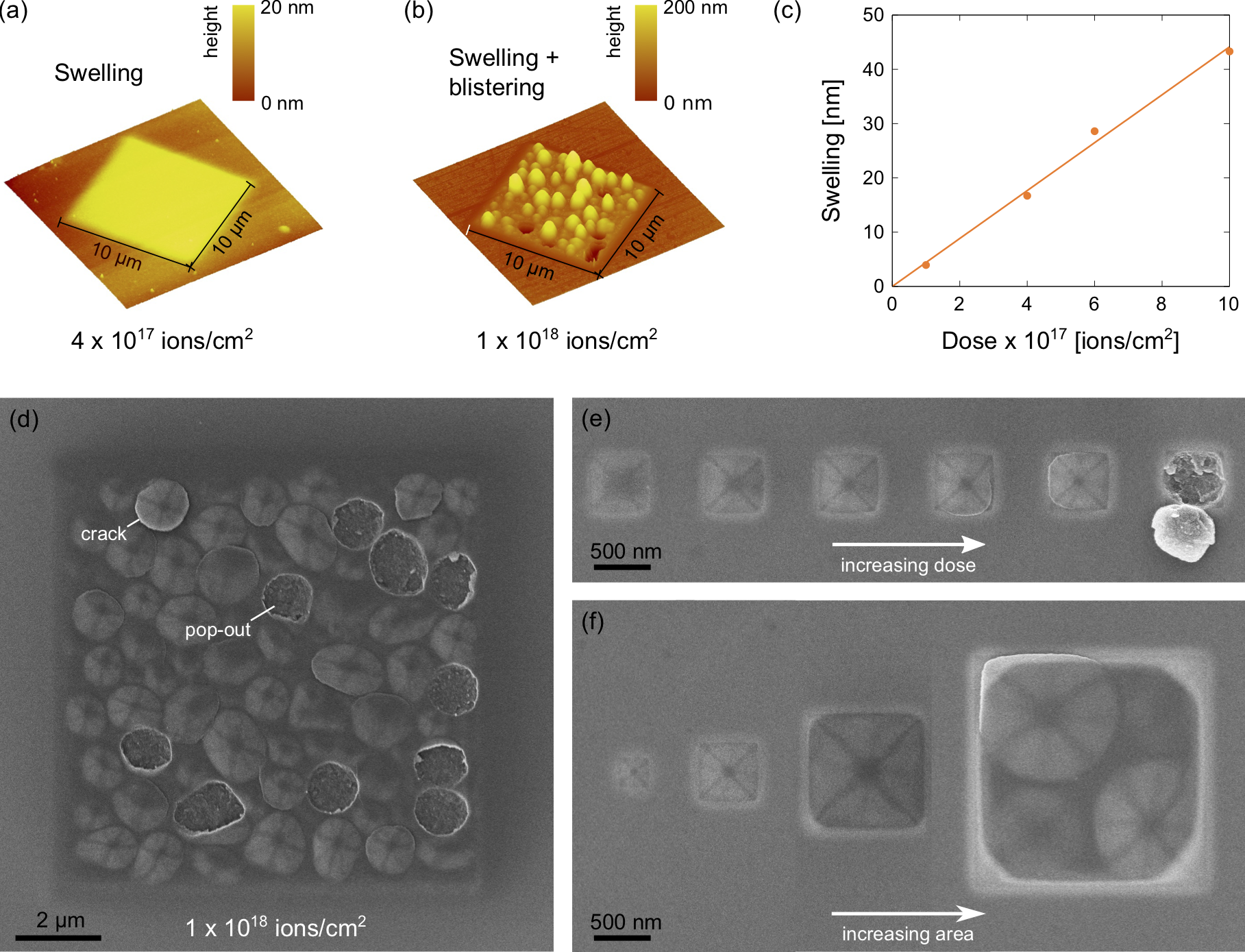}
  \caption{(a) and (b) AFM of W(100) sample irradiated with \unit[25]{keV} helium ions over an area of \unit[10$\times$10]{$\mu$m$^2$ to doses of \unit[4$\times$10$^{17}$ and 1$\times$10$^{18}$]{ions/cm$^2$}, respectively, showing surface swelling and eventual blistering. (c) Plot of surface swelling measured by AFM vs.\ irradiation dose with a linear fit. (d) HIM of \unit[10$\times$10]{$\mu$m$^2$} field irradiated with \unit[1$\times$10$^{18}$]{ions/cm$^2$}. (e) HIM showing a dose series for \unit[500$\times$500]{nm$^2$} fields irradiated with \unit[5$\times$10$^{17}$--1$\times$10$^{18}$]{ions/cm$^2$} in increments of \unit[1$\times$10$^{17}$]{ions/cm$^2$}.  (f) HIM of fields of increasing size (\unit[250$\times$250]{nm$^2$}, \unit[500$\times$500]{nm$^2$}, \unit[1$\times$1]{$\mu$m$^2$},
  \unit[2$\times$2]{$\mu$m$^2$}) each irradiated with \unit[7$\times$10$^{17}$]{ions/cm$^2$}.}}
  \label{Fig1}
\end{figure*}

In the following, we gain key experimental insights into the mechanism of helium-ion-induced blistering on tungsten using a Helium Ion Microscope for precise localized irradiation in combination with surface imaging by Helium Ion Microscopy (HIM) and Atomic Force Microscopy (AFM), and crucially, high-resolution imaging by Transmission Electron Microscopy (TEM) of cross-sections through isolated blisters prepared site-selectively by gallium focused ion beam (FIB) milling. 

The target used in this work is high-purity single-crystal tungsten in the form of a disk of diameter \unit[5]{mm} and height \unit[1]{mm}, with the (100) plane oriented parallel to the disk diameter. The top surface of the disk was mechanically polished to \unit[$\sim$1]{nm} root-mean-square surface roughness as verified by AFM (final step mechanical polish using a Buehler VibroMet). Helium ion implantation into the polished surface at room temperature was achieved by rastering the focused helium ion beam (He$^+$) of a Zeiss ORION NanoFab Helium Ion Microscope over selected areas \unit[250$\times$250]{nm$^2$} to \unit[10$\times$10]{$\mu$m$^2$} in size. The beam energy was set to \unit[25]{keV} and beam currents were selected by varying the aperture size: the smallest areas were irradiated at \unit[$\sim$1]{pA} (beam spot size \unit[0.5]{nm}) and the largest areas were irradiated using currents up to 100 pA (beam spot size tens of nanometers). The irradiation was programmed using NanoPatterning and Visualization Engine (NPVE) software from Fibics Inc., implementing a pixel dwell time of \unit[1]{$\mu$s} and a beam overlap between adjacent dwell points of \unit[50]{\%}. The irradiation doses ranged from \unit[1$\times$10$^{17}$--1$\times$10$^{18}$]{ions/cm$^2$}. 

Figures \ref{Fig1}(a) and (b) show AFM height maps measured ex-situ (Digital Instruments Nanoscope III) of \unit[10$\times$10]{$\mu$m$^2$} fields irradiated with \unit[4$\times$10$^{17}$ and 1$\times$10$^{18}$]{ions/cm$^2$}, respectively. At the lower dose a uniform rise in the irradiated area is observed and at the higher dose this swelling is accompanied by the formation of micron-sized blisters which may rupture or pop-out entirely. The threshold dose for blister formation was determined from dose series measurements to be \unit[$\sim$5$\times$10$^{17}$]{ions/cm$^2$}, in agreement with the critical dose for blister formation for tungsten and other metals measured elsewhere \cite{Chen2018,Erents1973}. Swelling height (neglecting local maxima and minima from blistering and pop-out events) increases linearly with dose, as shown in Fig.\ \ref{Fig1}(c). 

\begin{figure*}
    \centering
        \includegraphics[scale=.9]{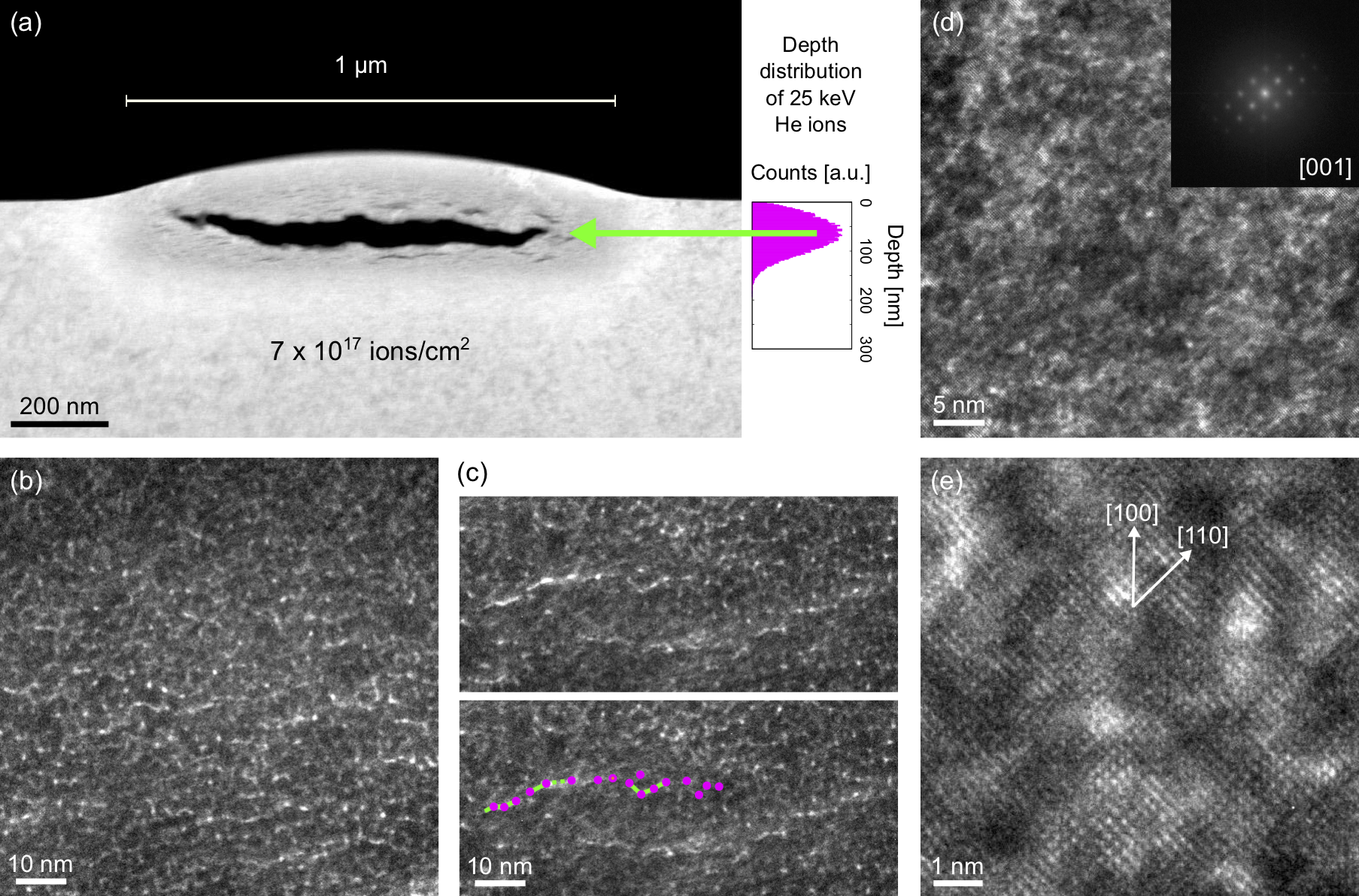}
  \caption{(S)TEM results for a cross-section through a blister in the W(100) sample formed by irradiating a \unit[1$\times$1]{$\mu$m$^2$} field with \unit[25]{keV} helium ions to a dose of \unit[7$\times$10$^{17}$]{ions/cm$^2$}. (a) HAADF-STEM of the blister showing the main cavity beneath the shell and nanocracks above and below the cavity; the helium depth distribution calculated by SRIM is shown plotted to scale for comparison. (b) Bright-field TEM of a region near the center of the bliser shell showing helium nanobubbles. (c) TEM of a region nearer the edge of the shell and its annotated duplicate with the onset of nanocrack formation indicated. (d) and (e) High-resolution TEM showing nanobubbles and host tungsten crystal lattice; FFT inset of (d) indexes to the [001] zone axis in agreement with the known orientation of the target.}
  \label{Fig2}
\end{figure*}

HIM micrographs of irradiated fields (measured in-situ) are shown in Figs.\ \ref{Fig1}(d)--(f). The HIM images are obtained by detecting the secondary electrons generated by the interaction of the scanning ion beam with the sample; since the escape depth of the secondary electrons is only a few nanometers, it is primarily the surface that is probed. In Fig.\ \ref{Fig1}(d) we see a HIM image of a \unit[10$\times$10]{$\mu$m$^2$} field that was irradiated to the higher dose of \unit[1$\times$10$^{18}$]{ions/cm$^2$}. An array of blisters is observed, some of which have popped out, as also seen in the AFM height map of Fig.\ \ref{Fig1}(b). In addition, a few of the remaining blisters show bright contrast along their perimeters, presumably indicative of the crack that will eventually result in pop-out. A characteristic feature of the HIM images is the appearance of dark bands forming cross shapes on the blister shells, as has been noted in HIM work performed elsewhere \cite{Veligura2013,Hlawacek2016}. This effect is caused by the dome shape of the blister shell, which locally either blocks or allows access to certain lattice planes for ion channeling. Channeling ions travel deeper into the sample generating fewer secondary electrons near the surface thus giving rise to darker contrast. Here we observe two cross shapes on each blister, one rotated by \unit[45]{$^\circ$} with respect to the other, which we attribute to channeling along (100) and (110) planes \cite{Nordlund2016}. The dose series in Fig.\ \ref{Fig2}(e) for \unit[500$\times$500]{nm$^2$} fields irradiated with \unit[5$\times$10$^{17}$--1$\times$10$^{18}$]{ions/cm$^2$} tracks blister growth and development of the channeling conditions for a single blister until finally the blister pops out. In this example, pop-out occurred at the \unit[1$\times$10$^{18}$]{ions/cm$^2$} dose and the inverted blister shell is observed still partially attached allowing inspection of each ruptured surface. In order to probe the preferred blister size, areas of increasing size from \unit[250$\times$250]{nm$^2$} to \unit[2$\times$2]{$\mu$m$^2$} were irradiated with \unit[7$\times$10$^{17}$]{ions/cm$^2$}, as shown in Fig.\ \ref{Fig1}(f). Fields of size \unit[1$\times$1]{$\mu$m$^2$} and below only developed a single blister, whereas four blisters occupied the \unit[2$\times$2]{$\mu$m$^2$} field, in accordance with the preferred blister size of \unit[$\sim$1]{$\mu$m$^2$} observed for the larger irradiated fields. We note that the HIM imaging doses were at least three orders of magnitude lower than those used to induce the swelling and blistering, and for regions to be subsequently investigated by AFM or TEM, doses of at least six orders of magnitude lower were implemented.

High-resolution analysis of structural changes below the surface was achieved by TEM and scanning TEM (STEM) of \unit[$\sim$100]{nm} thick cross-sections through blisters prepared following the gallium FIB lift-out method \cite{Mayer2007} (FEI Helios G4 UX DualBeam). In order to shield the blister surface during the FIB milling steps a protective cap was first deposited onto the blister by focused electron-beam-induced deposition using a platinum-based precursor. Figure \ref{Fig2} shows a set of (S)TEM results (FEI TitanX operated at \unit[300]{kV}) for a cross section through a blister formed using a dose of \unit[7$\times$10$^{17}$]{ions/cm$^2$} irradiated over an area of \unit[1$\times$1]{$\mu$m$^2$}. The high-angle annular dark-field (HAADF) STEM image in Fig.\ \ref{Fig2}(a) clearly shows the cavity that forms under the blister shell, which is responsible for the dome-shaped blister surface. In addition, dark lines (referred to hereafter as nanocracks) are observed above and below the main cavity, which we note were not discernible in bright-field TEM images acquired at the same magnification.

Simulations using the Monte Carlo Stopping and Range of Ions in Matter (SRIM) code \cite{Ziegler2010} for \unit[25]{keV} helium ions incident on a tungsten target give a depth distribution with an intensity maximum at \unit[67]{nm}. We plot this distribution to scale to the right of the STEM image of Fig.\ \ref{Fig2}(a), with the origin of the plot aligned to the height of the tungsten surface outside the irradiated region. While the simulations assume an amorphous target and hence neglect channeling and do not account for the gradual decrease in material density due to helium implantation (both of these effects resulting in larger penetration depths), the peak in the plot lines up with the location of the blister cavity remarkably well, as shown by the green arrow. The locations of the nanocracks also fall within the envelope of the helium depth distribution, with nanocracks observed both above (i.e.\ in the blister shell) and below the main cavity, causing swelling in both regions. These are key observations, since as summarized below there has been much debate over the discrepancy between the position of the helium peak and the thickness of the blister shell, which our new experimental approach is able to directly address.

Early investigations of blistering indeed presumed that blister formation must be initiated at a depth corresponding to the so-called helium peak, which led to the proposed ``gas pressure model" based on the internal release of the implanted helium into a disk-shaped cavity that nucleated parallel to the surface by interbubble coalescence/fracture \cite{Evans1977,Yadava1981}. However, subsequent measurements of blister shells revealed shell thicknesses that were in fact greater than the depth of the helium peak, which cast doubt on the gas-driven model. An alternative ``lateral stress model" was proposed, based on stress buildup at a depth corresponding to the maximum stopping distance of the helium ions (i.e.\ at the interface between the layer of material containing helium nanobubbles and the pristine material below). Yet the stress model alone could not explain the spherical blister shape, instead favoring the development of a rippled surface. Thus  interest in explaining the blister shell thickness in the framework of a gas-driven model persisted and in fact already early on, swelling due to the presence of helium nanobubbles in the blister shell to account for the observed increase in shell thickness was proposed \cite{Terreault1977}. With the present work we definitively show that nanocracks in the blister shell are responsible for its increased thickness and by creating an isolated blister for cross-sectional analysis, a true comparison of the depth of the crack plane of the main blister cavity with respect to the adjacent non-irradiated surface is made possible, confirming that the location of the blister cavity in the tungsten target does indeed correlate with the helium peak.

Upon inspection of the blister shell at higher magnification using bright-field TEM, helium nanobubbles of diameter \unit[$\sim$1]{nm} are revealed, as seen in Fig.\ \ref{Fig2}(b). These bright spots (imaged slightly underfocus in order to enhance contrast) resemble the helium nanobubbles observed by TEM in irradiated samples elsewhere \cite{Johnson1999,Donnelly2014}. (As a control, unexposed regions were imaged and in those areas the bright spots corresponding to nanobubbles were not observed.) In some regions connectivity between nanobubbles is detected, appearing to favor a horizontal direction, i.e.\ parallel to the target surface, as highlighted in Fig.\ \ref{Fig2}(c). To the best of our knowledge, this observation provides the first direct experimental evidence for the ``interbubble fracture model" put forward by Evans \cite{Evans1977,Evans1978}, in which layers of pressurized helium bubbles are proposed to coalesce to form cracks, the local accumulation thereof eventually forming the blister cavity. By imaging at higher resolution to resolve the lattice planes of the crystal we are also able to confirm that despite the presence of the helium nanobubbles, the crystallinity of the host lattice is preserved (see high-resolution TEM results in Figs.\ \ref{Fig2}(d) and (f)).  

In summary, by using the HIM approach to selectively form isolated blisters in tungsten at various stages of development, cross-sectioning by FIB milling and imaging by (S)TEM, we conclude that a) the blister cavity does indeed form at the depth of the helium peak, b) the thickness increase of the blister shell is primarily caused by swelling due to the nanocracks that form throughout the implanted volume, and c) the nanocracks (and ultimately the main blister cavity) form as a result of lateral coalescence of the nanobubbles. We note that while the present work confirms the validity of the gas pressure model for tungsten, which is a low fracture toughness metal, additional factors likely play a role in the blistering of ductile metals such as copper \cite{Johnson1999}. Surface blistering mechanisms for a range of materials can now be closely scrutinized following the experimental approach described herein. 

\section*{Acknowledgements}

This work was performed at the Biomolecular Nanotechnology Center, a core facility of the California Institute for Quantitative Biosciences, and at the Molecular Foundry, \newline Lawrence Berkeley National Laboratory, which is supported by the Office of Science, Office of Basic Energy Sciences, of the U.S. Department of Energy under Contract No.\ DE-AC02-05CH11231. The authors also acknowledge funding from NSF DMR Award No.\ 1807822.

\bibliographystyle{elsarticle-num}

\bibliography{bib}

\providecommand{\BIBKh}{Kh}
\begin{thebibliography}{10}
\expandafter\ifx\csname url\endcsname\relax
  \def\url#1{\texttt{#1}}\fi
\expandafter\ifx\csname urlprefix\endcsname\relax\def\urlprefix{URL }\fi
\expandafter\ifx\csname href\endcsname\relax
  \def\href#1#2{#2} \def\path#1{#1}\fi

\bibitem{Bolt2002}
H.~Bolt, V.~Barabash, G.~Federici, J.~Linke, A.~Loarte, J.~Roth, K.~Sato, J.
  Nucl. Mater. 307--311~(1) (2002) 43--52.

\bibitem{Nelson1964}
R.~S. Nelson, Philos. Mag. 9~(98) (1964) 343.

\bibitem{Primak1966}
W.~Primak, J.~Luthra, J. Appl. Phys. 37~(6) (1966) 2287.

\bibitem{Wang2016}
Z.-Y. Wang, F.~I. Allen, Z.-W. Shan, P.~Hosemann, Acta Mater. 121~(9) (2016)
  78.

\bibitem{Mairov2019}
A.~Mairov, D.~Frazer, P.~Hosemann, K.~Shridharan, Scr. Mater. 162 (2019) 156.

\bibitem{Shan2019}
G.~B. Shan, Y.~Z. Chen, N.~N. Liang, H.~Dong, W.~X. Zhang, T.~Suo, F.~Liu,
  Mater. Lett. 238 (2019) 261.

\bibitem{Veligura2013}
V.~Veligura, G.~Hlawacek, R.~P. Berkelaar, R.~van Gastel, H.~J.~W. Zandvliet,
  B.~Poelsema, Beilstein J. Nanotechnol. 4 (2013) 453.

\bibitem{Bergner2018}
F.~Bergner, G.~Hlawacek, C.~Heintze, J. Nucl. Mater. 505 (2018) 267.

\bibitem{Chen2018}
Z.~Chen, L.-L. Niu, Z.~Wang, L.~Tian, L.~Kecskes, K.~Zhu, Q.~Wei, Acta Mater.
  147~(1) (2018) 100.

\bibitem{Alimov2013}
V.~{\BIBKh}. Alimov, Y.~Hatano, B.~Tyburska-P{\"u}schel, K.~Sugiyama,
  I.~Takagi, Y.~Furuta, J.~Dorner, M.~Fu{\ss}eder, K.~Isobe, T.~Yamanishi,
  M.~Matsuyama, J. Nucl. Mater. 441~(1--3) (2013) 280--285.

\bibitem{Cipiti2005}
B.~B. Cipiti, G.~L. Kulcinski, J. Nucl. Mater. 347~(3) (2005) 298.

\bibitem{Petty2015}
T.~J. Petty, M.~J. Baldwin, M.~I. Hasan, R.~P. Doerner, J.~W. Bradley, Nucl.
  Fusion 55~(9) (2015) 093033.

\bibitem{Parish2017}
C.~M. Parish, K.~Wang, R.~P. Doerner, M.~J. Baldwin, Scr. Mater. 127 (2017)
  132.

\bibitem{Evans1977}
J.~H. Evans, J. Nucl. Mater. 68 (1977) 129.

\bibitem{Evans1978}
J.~H. Evans, J. Nucl. Mater. 76 \& 77 (1978) 228.

\bibitem{Erents1973}
S.~K. Erents, G.~M. McCracken, Radiat. Eff. 18~(3-4) (1973) 191.

\bibitem{Hlawacek2016}
G.~Hlawacek, V.~Veligura, R.~van Gastel, B.~Poelsema, Helium Ion Microscopy,
  Springer International Publishing Switzerland, 2016, Ch.~9.

\bibitem{Nordlund2016}
K.~Nordlund, F.~Djurabekova, G.~Hobler, Phys. Rev. B 94 (2016) 214109.

\bibitem{Mayer2007}
J.~Mayer, L.~A. Giannuzzi, T.~Kamino, J.~Michael, MRS Bulletin 32~(5) (2007)
  400.

\bibitem{Ziegler2010}
J.~F. Ziegler, M.~D. Ziegler, J.~P. Biersack, Nucl. Instrum. Methods Phys. Res.
  B 268~(11--12) (2010) 1818.

\bibitem{Yadava1981}
R.~D.~S. Yadava, J. Nucl. Mater. 98~(47).

\bibitem{Terreault1977}
B.~Terreault, J.~G. Martel, R.~G. St.-Jacques, G.~Veilleux, J.~L'Ecuyer,
  C.~Brassard, C.~Cardinal, L.~Deschdnes, J.~Labrie, J. Nucl. Mater. 68~(3)
  (1977) 334.

\bibitem{Johnson1999}
P.~B. Johnson, R.~W. Thomson, K.~Reader, J. Nucl. Mater. 273 (1999) 117.

\bibitem{Donnelly2014}
S.~E. {Donnelly}, in: 2014 International Conference on Manipulation,
  Manufacturing and Measurement on the Nanoscale (3M-NANO), 2014, pp. 18--22.
\newblock \href {http://dx.doi.org/10.1109/3M-NANO.2014.7057355}
  {\path{doi:10.1109/3M-NANO.2014.7057355}}.

\end{thebibliography}

\end{document}